\documentclass[preprint,floats,aps,12pt,superscriptaddress]{revtex4}
\usepackage{natbib} 
\usepackage{times} 
\usepackage{amssymb,amsmath}
\usepackage{prelim2e}\usepackage[none,bottom]{draftcopy}
\draftcopyName{preprint / }{1.2} %ADAPT TEXT
\ifx\pdfoutput\undefined
\usepackage[usenames,dvips]{color}
\usepackage{graphicx}     % Include figure files
\else
\usepackage[usenames,dvipsnames]{color}
\usepackage{epstopdf}
\usepackage[pdftex]{graphicx}
\usepackage[pdftex]{hyperref}
\hypersetup{a4paper,
    pdftitle={Manuscript on XXX} %ADAPT TEXT 
    pdfauthor={et al. and Uwe Thiele},  %ADAPT TEXT 
pdfproducer={lateX},
pdfview=FitV,       % FitH
pdfstartview=FitB,
linkcolor=blue,     % links to same page
citecolor=blue,     % citations
pagecolor=blue,     % links to same page
urlcolor=red,      % links to URLs
breaklinks=true,    % links may be split onto 2 lines
colorlinks=true,
citebordercolor=0 0 0,  % color for \cite
filebordercolor=0 0 0,
linkbordercolor=0 0 0,
menubordercolor=0 0 0,
pagebordercolor=0 0 0,
urlbordercolor=0 0 0,
pdfhighlight=/I,
pdfborder=0 0 0,   % no box around links
backref=false,
pagebackref=false,
bookmarks=true,
bookmarksopen=true,
bookmarksnumbered=true
}
\fi
\ifx\pdfoutput\undefined
\DeclareGraphicsExtensions{.eps}
\else
\DeclareGraphicsExtensions{.jpg, .pdf, .tif, .png}
\fi
\graphicspath{{.},{./Fig/}} %$ %ADAPT DIRECTORY
\usepackage[usenames,dvipsnames]{color}
\usepackage[normalem]{ulem}
\renewcommand{\vec}[1]{\mathbf{#1}}
\newcommand{\vecg}[1]{\boldsymbol{#1}}
\newcommand{\tens}[1]{\mathbf{\underline{#1}}}

\begin{document}
%
%----------------------------------------------------------------%
\title{Recent advances in and future challenges for mesoscopic hydrodynamic modelling of complex wetting}
\author{Uwe Thiele}
\email{u.thiele@uni-muenster.de}
\homepage{http://www.uwethiele.de}
\affiliation{Institut f\"ur Theoretische Physik, Westf\"alische Wilhelms-Universit\"at M\"unster, Wilhelm Klemm Str.\ 9, 48149 M\"unster, Germany}
\affiliation{Center of Nonlinear Science (CeNoS), Westf{\"a}lische Wilhelms-Universit\"at M\"unster, Corrensstr.\ 2, 48149 M\"unster, Germany}
\affiliation{Center for Multiscale Theory and Computation (CMTC), Westf{\"a}lische Wilhelms-Universit\"at M\"unster, Corrensstr.\ 40, 48149 M\"unster, Germany}
\begin{abstract}
We highlight some recent developments that widen the scope and reach of mesoscopic thin-film (or long-wave) hydrodynamic models employed to describe the dynamics of thin films, drops and contact lines of simple and complex liquids on solid substrates. The basis of the discussed developments is the reformulation of various mesoscopic thin-film hydrodynamic models  as gradient dynamics on underlying energy functionals. After briefly presenting the general approach, the following sections discuss how to improve these models by amending the energy functional and the mobility function, how to obtain gradient dynamics models for some complex liquids, and how to incorporate processes beyond relaxational dynamics.\\[1ex]
\end{abstract}
%
%\begin{keyword} 
%Sliding drops \sep Heterogeneous substrates \sep Pinning and depinning
%\pacs{
%68.15.+e, % Thin films: Liquid thin films
%47.20.Ky  % Fluid dynamics: Nonlinearity (including bifurcation theory)
%47.55.Dz  % Drops and bubbles 
%68.08.-p  % Liquid-solid interfaces
%}
%\end{keyword} 
%
\maketitle
%
%\received{6.5.2002}
%
%----------------------------------------------------------------%
%
%%%%%%%%%%%%%%%%%%%%%%%%%%%%%%%%%%%%%%%%%%%%%%%%%%%%%%%%%%
%  INTRO
%%%%%%%%%%%%%%%%%%%%%%%%%%%%%%%%%%%%%%%%%%%%%%%%%%%%%%%%%%
%%%%%%%%%%%%%%%%%%%%%%%%%%%%%%%%%%%%%%%%%%%%%%%%%%%%%%%%%%%%%%%%%%%%%%%%%%%%%%%
\section{Introduction} \label{sec:intro}
%%%%%%%%%%%%%%%%%%%%%%%%%%%%%%%%%%%%%%%%%%%%%%%%%%%%%%%%%%%%%%%%%%%%%%%%%%%%%%%
%
For many years, the wider field of dynamic wetting and dewetting is a very active and frequently reviewed field of research. General reviews were published, for instance, 1985 by de Gennes \cite{Genn1985rmp}, 1992 by Leger and Joanny \cite{LeJo1992rpp}, 2009 by Bonn et al.\ \cite{BEIM2009rmp} and by Starov and Velarde \cite{StVe2009jpm}. These works review static (equilibrium) wetting behaviour as well as dynamic (nonequilibrium) behaviour and focus on experimental finding as well as on theoretical descriptions on the macroscopic scale (based on interface tensions) or on the mesoscopic scale (incorporating a wetting potential).  Related reviews focus on wetting hydrodynamics \cite{TeDS1988rpap}, hydrodynamic long-wave models \cite{OrDB1997rmp,Thie2007,CrMa2009rmp}, the motion of the contact line \cite{Duss1979arfm,SnAn2013arfm}, the role of surfactants in thin-film flows \cite{MaCr2009sm}, molecular dynamics simulations of wetting \cite{DeBl2008armr} and wetting in complex geometries \cite{HeBS2008armr}.
%similar equations in biopysics \cite{GaCo1996hcr}

Here, we do not present a general review of the whole field or of particular phenomena related to dynamic wetting. Instead, the present contribution solely aims at highlighting a number of recent developments that concern the theoretical description of wetting and dewetting dynamics via mesoscopic (or long-wave) hydrodynamic models. They are frequently employed to describe the dynamics of thin films, drops and contact lines of simple and complex liquids on solid substrates. Reformulating the various mesoscopic thin-film hydrodynamic models as gradient dynamics on underlying energy functionals forms the basis of these recent developments that widen the scope and reach of mesoscopic thin-film hydrodynamics.

The here-considered thin-film equation is in its basic form a mesoscopic hydrodynamic description of the time evolution of height profiles
$h(x,y,t)$ of films and drops of simple partially wetting liquids on flat solid substrates (corresponding to the $(x,y)$-coordinate plane):
\begin{equation}
\partial_t h \,=\,-\nabla\cdot\left[Q(h)\nabla(\gamma\Delta h + \Pi(h))\right]
\label{eq:film}
\end{equation}
where $Q(h)=h^3/3\eta$ is the mobility function in the case without slip at the substrate, $-\gamma\Delta h$ is the Laplace pressure representing capillarity, and $\Pi(h)$ is the Derjaguin (or disjoining) pressure that models wettability in mesoscopic hydrodynamic models. The parameters $\eta$ and $\gamma$ are the dynamic viscosity of the liquid and the liquid-gas interface tension, respectively. Here, $\partial_t$ denotes the partial time derivative, $\nabla=(\partial_x,\partial_y)^T$ is the two-dimensional (2d) gradient operator and $\Delta=\nabla\cdot \nabla$ the corresponding laplacian.

It was noted by Mitlin \cite{Mitl1993jcis} that the thin-film equation (\ref{eq:film}) can be brought into the form of a gradient dynamics of a conserved order parameter field, i.e.,
\begin{equation}
\partial_t h \,=\,
\nabla\cdot\left[Q(h)\nabla\frac{\delta
    \mathcal{F}}{\delta h}\right].
\label{eq:onefield:gov}
\end{equation}
That is, the field $h$ follows a mass-conserving dynamics where mass refers to $m=\int hdxdy$. 
Here, $\mathcal{F}[h]$ is an energy functional and $\delta/\delta h$ denotes the variational derivative.  

Other prominent examples of such gradient dynamics models are the Cahn-Hilliard equation describing the decomposition of a binary mixture \cite{Cahn1965jcp}, evolution equations for surface profiles in epitaxial growth \cite{GoDN1999pre} and dynamical density functional theories (DDFT) for the dynamics of colloidal particles \cite{MaTa1999jcp,ArRa2004jpag}.  For volatile liquids, Eq.~\eqref{eq:onefield:gov} may be extended to contain a conserved and a nonconserved contribution to the dynamics \cite{Thie2010jpcm}. The form of such conserved and nonconserved gradient dynamics models can be derived employing Onsager's variational principle \cite{Doi2011jpcm}. This is shown in the appendix~\ref{sec:grad-dyn}.

For systems dominated by capillarity and wettability the functional in long-wave approximation is
\begin{equation}
\mathcal{F}[h]\,=\,\int\left[\frac{\gamma}{2}(\nabla h)^2 + f(h) \right]d^2r
\label{eq:en1}
\end{equation}
and is sometimes called ``interface hamiltonian'' \cite{Diet1988,schi1990,MuMa2003jpcm}. It contains a squared-gradient contribution that results in the Laplace pressure term of Eq.~\eqref{eq:film}, and the wetting energy (or binding potential) $f(h)$ that results via $\Pi=-df/dh$ in the Derjaguin pressure term of Eq.~\eqref{eq:film}. The wetting potential acts on the mesoscopic length scale, i.e., $f(h)\to0$ for $h\to\infty$. For partially wetting liquids it normally has a minimum at the height $h_\mathrm{a}$ of an adsorption layer that coexists with macroscopic drops of equilibrium contact angle $\theta_\mathrm{e}=\sqrt{-f(h_\mathrm{a})/\gamma}$. The consistency condition of the macroscopic Young-Dupr\'e law ($\gamma\cos\theta_\mathrm{e} = \gamma_\mathrm{sg} - \gamma_\mathrm{sl}$) and the mesoscopic description is $f(h_\mathrm{a})=\gamma_\mathrm{sg} -\gamma_\mathrm{sl} - \gamma \equiv S$ where $S$ is the spreading coefficient and $\gamma_\mathrm{sg}$ and $\gamma_\mathrm{sl}$ are the solid-gas and solid-liquid interface tension, respectively. For more details see \cite{ChSD1982jcis,Genn1985rmp,DoIn1993pa,GennesBrochard-WyartQuere2004,BEIM2009rmp,StVe2009jpm,TSTJ2018l}.

Here we review some of the important steps taken the past five years that widen the scope and reach of mesoscopic thin-film hydrodynamics in several ways and are all based in the reformulation of thin-film model as gradient dynamics.

In particular, the following sections briefly discuss how to extract wetting potentials $f(h)$ from microscopic models as Molecular Dynamics simulations and density functional theories (section~\ref{sec:avd-wetten}); how to use the gradient dynamics form (\ref{eq:onefield:gov}) as a guide to obtain better models without employing an asymptotic approach (section~\ref{sec:avd-curvature}); how to extend the mobility $Q(h)$ in such a way that the model goes beyond hydrodynamics (section~\ref{sec:avd-mobility}); how to obtain gradient dynamics models for some complex liquids (section~\ref{sec:avd-grad}); and how to use thermodynamically consistent gradient dynamics models as the base of models for biofilms (section~\ref{sec:avd-biofilm}).
%and how to coarse-grain thin-film models further, e.g., to obtain dynamic models for the statistics of drop ensembles (section~\ref{sec:avd-stat}). 
Section~\ref{sec:conc} concludes and gives a brief outlook.

%%%%%%%%%%%%%%%%%%%%%%%%%%%%%%%%%%%%%%%%%%%%%%%%%%%%%%%%%%%%%%%%%%%%%%%%%%%%%%%
\section{Obtaining wetting potentials from microscopic models}\label{sec:avd-wetten}
%%%%%%%%%%%%%%%%%%%%%%%%%%%%%%%%%%%%%%%%%%%%%%%%%%%%%%%%%%%%%%%%%%%%%%%%%%%%%%%
%
To practically use Eq.~\eqref{eq:onefield:gov} with an energy \eqref{eq:en1} in an analysis of the dynamics of dewetting and spreading or of the motion of sliding drops one needs to employ a particular wetting energy $f(h)$ (or Derjaguin pressure $\Pi(h)$). Usual choices are combinations of power laws \cite{Mitl1993jcis,BeGW2001n,BGSM2003nm,PiTh2006pf,KiRW2011n}, combinations of exponentials \cite{PiPo2000pre,BeNe2001prl,TVNB2001pre} or the combination of a power law and an exponential \cite{ShJa1993jcis,ShKh1998prl,ThVN2001prl,BKHT2011pre}. Typical expressions are reviewed in \cite{TeDS1988rpap,OrDB1997rmp}.  In principle, all these forms for $f(h)$ are asymptotic expressions that are only strictly valid for large $h$ and should therefore be expected to break down as $h \to 0$. For instance, all expressions involving power laws show an unphysical divergence for $h\to0$ that is sometimes avoided by introducing a lower cut-off thickness \cite{Genn1985rmp}.

However, alternatively to such combinations of different asymptotic expressions it has recently been shown that one may directly extract wetting energies (and interfacial tensions) from various microscopic models, namely, Molecular Dynamics (MD) simulations \cite{TMTT2013jcp}, lattice Density Functional Theory (DFT) \cite{HuTA2015jcp,BTAH2017jcp} and continuous DFT \cite{HuTA2017jcp} for Lennard-Jones fluids and other simple liquids. Refs.~\cite{MaMu2006jcp,MacD2011epjst,MBKP2014acis} also extract wetting potentials from computer simulations.

The continuous DFT used in \cite{HuTA2017jcp} is based on fundamental measure theory \cite{HansenMcDonald2006} and therefore includes the influence of the layering of molecules close to interfaces and the resulting oscillatory density profiles. The extracted wetting potentials are represented by well defined fit functions that can easily be incorporated in mesoscale hydrodynamic models \cite{YSTA2017pre}. The calculated static drop shapes are shown to be in quite good agreement with drop profiles determined directly from the microscopic DFT \cite{HuTA2017jcp}. There, for some systems oscillatory wetting potentials are found that may result in pronounced terraced drop profiles \cite{YSTA2017pre}.

It is to note that some details of the extraction of wetting potentials are tricky and there remain some open questions, see, for instance, the discussion of the ``fictitious potential'' that in the DFT approach is needed to stabilize particular film heights at imposed coexistence chemical potential \cite{HuTA2015jcp,HuTA2017jcp} and the limitations encountered for MD simulations at unstable film heights \cite{TMTT2013jcp}. However, results obtained with the fictitious potential approach \cite{HuTA2015jcp} and a nudged elastic band approach \cite{BTAH2017jcp} do very well agree giving further support to both approaches.

The obtained wetting energies do not show a divergence for $h\to0$ and can therefore be employed to study drops on truly dry substrates. This will then also allow one to distinguish in mesoscale simulations the dry case (drop on dry substrate) and the moist case (drop on adsorption layer) discussed in \cite{Genn1985rmp}.

At the moment, the missing element is the extraction of transport coefficients from microscopic simulations (as MD or kinetic Monte Carlo (MC) simulations) to come to a fully quantitative mapping of microscopic and mesoscopic theories not only for the static but also for the dynamic behaviour. Qualitative comparisons of transitions in the dynamic behaviour exist, e.g., between kinetic MC simulations and thin-film modelling of the Plateau-Rayleigh instability of liquid ridges \cite{TBHT2017jcp}. To reach a quantitative agreement one would need to measure diffusive and convective mobilities in microscopic dynamic models to introduce them into mesoscale gradient dynamics models (see section~\ref{sec:avd-mobility}). The described concepts have to our knowledge not yet been applied to complex liquids.

%%%%%%%%%%%%%%%%%%%%%%%%%%%%%%%%%%%%%%%%%%%%%%%%%%%%%%%%%%%%%%%%%%%%%%%%%%%%%%%
\section{Improving the energy functional}\label{sec:avd-curvature}
%%%%%%%%%%%%%%%%%%%%%%%%%%%%%%%%%%%%%%%%%%%%%%%%%%%%%%%%%%%%%%%%%%%%%%%%%%%%%%%
%
It is known for some time that long-wave thin-film models can sometimes be improved by modifying the curvature terms. This was first introduced in Ref.~\cite{GaRa1988ces} for the break-up of liquid films in cylindrical capillaries and is also used in models of airway closure \cite{HeWh2002jfm}. A similar improvement was introduced as small-curvature approximation (but allowing for large surface slopes) in Ref.~\cite{Snoe2006pf} for situations involving moving contact lines (also used in \cite{SADF2007jfm,GLFD2016jfm}).  Also in the context of surface waves on a falling film, a regularised Kuramoto-Sivashinsky (KS) equation has been analysed where the long-wave curvature is replaced by the exact curvature \cite{BKOR1992pla}.
In the light of the gradient dynamics viewpoint advocated here, we suggest that the success of these modifications results from the improved representation of the underlying energy functional that seems to be more important for a correct description of the physical behaviour than the details of the dynamics. With other words it is more important to improve $\mathcal{F}$ in Eq,~(\ref{eq:onefield:gov}) than the mobility $Q$.

The question was recently investigated in detail in Ref.~\cite{BoTH2018jfm} for a particular nonequilibrium system, namely, a liquid film on the outside of a rotating cylinder \cite{Moff1977jdm,DuWi1999jfm,LRTT2016pf} that represents a typical coating flow \cite{WeRu2004arfm}. Quite a number of different long-wave models exist for this system \cite{Pukh1977pmtf,Kelm2009jfm,Thie2011jfm}.
Considering the overdamped limit (no inertia), Ref.~\cite{BoTH2018jfm} computes steady solutions of the free-surface Stokes equations using a moving-mesh, finite-element method combined with a spatially adaptive pseudo-arclength continuation method \cite{DWCD2014ccp}. The results are compared with (i) previously employed leading order and next order asymptotically correct long-wave thin-film models and (ii) a variational thin-film model. The latter is based on the reformulation of the leading order thin-film model as gradient dynamics of the form (\ref{eq:onefield:gov}). The mobility function and energy are as obtained from the long-wave approximation and an additional nonvariational lateral driving term (of the form of a comoving-frame term) results from the cylinder rotation.  Then, however, the long-wave energy functional (surface and potential energy) is replaced by the exact one, e.g., the first term of the energy (\ref{eq:en1}) is replaced by $\gamma\sqrt{1+(\nabla h)^2}$.

The results are rather striking: Fixing the liquid volume on the cylinder and decreasing the speed of cylinder rotation starting from a large value, employing the Stokes equations one finds a saddle-node bifurcation at some critical rotation speed. At smaller speed, no steady solutions are found because the liquid drips off the cylinder. However, no such limiting point is found employing the leading order and next order long-wave models that both give a family of steady pending drop profiles down to zero rotation speed. In stark contrast, the variational thin-film model qualitatively agrees with the Stokes equation and also gives a much better quantitative agreement than the other models in all the considered parameter region.  Note, that for liquid films inside a rotating cylinder (rimming flow) all models qualitatively agree \cite{BoTH2018jfm}.

The advantage of the gradient dynamics viewpoint is that it allows us to rationalise the known ``exact-curvature trick'' as an example of a well defined strategy to improve dynamical models.  It consists in making the energy as exact as possible and keeping the mobility function (or transport coefficients) as simple as possible.  This ensures the correct representation of static equilibria and only approximates the dynamics. The approach may be seen as not being `rational' in an asymptotic (long-wave approximation) sense because part of the neglected terms are of the same order in the smallness parameter~\footnote{Here given by the ratio of length scales
  perpendicular and parallel to the substrate.} as the smallest kept terms. We argue, however, that the gradient dynamics form (\ref{eq:onefield:gov}) is a highly important feature that should be conserved in the approximation process as the Stokes equation itself can be written as a gradient dynamics, i.e., it may be obtained via Onsager's variational principle \cite{Doi2013}. Actually, another prominent example where such a strategy is followed is the Cahn-Hilliard equation describing the decomposition of a binary mixture \cite{CaHi1958jcp,Cahn1965jcp}. There, the energy is of the form (\ref{eq:en1}) with $f$ being approximated by a fourth order polynomial while the mobility function $Q$ in (\ref{eq:onefield:gov}) is simply assumed to be a constant. A version with better approximations of both, energy and mobility, can be found, e.g., in \cite{KoOt2002cmp}, however, only the energy is improved on in \cite{MoYO2001jpsj,ChMZ2011mjm}.

In the future, the presented basic idea of employing the gradient dynamics form of thin-film (long-wave or lubrication) equations as starting points for model improvements should be systematically tested for more systems. Namely, existing long-wave models should first be brought in gradient dynamics form as, e.g., done for two-layer films in Refs.~\cite{PBMT2004pre,JHKP2013sjam}. Then a better approximation or even the exact form of the underlying energy functional can be employed (keeping the same mobility functions) to obtain an improved model. For instance, for the two-layer films the exact curvatures could be used instead of the long-wave ones.

Section~\ref{sec:avd-grad} below discusses how long-wave models for some complex fluids (solutions, suspensions, mixtures) are brought into gradient dynamics form opening an avenue for many model extensions that are automatically thermodynamically consistent.  An important point that also needs more consideration is the question which nongradient driving terms (for some examples and a tentative classification for scalar order parameter fields see \cite{EGUW2018springer}) can be added to the general gradient dynamics model without changing the dominance of the energy functional discussed in this section.
%
%%%%%%%%%%%%%%%%%%%%%%%%%%%%%%%%%%%%%%%%%%%%%%%%%%%%%%%%%%%%%%%%%%%%%%%%%%%%%%%
\section{Mobility for various transport channels}\label{sec:avd-mobility}
%%%%%%%%%%%%%%%%%%%%%%%%%%%%%%%%%%%%%%%%%%%%%%%%%%%%%%%%%%%%%%%%%%%%%%%%%%%%%%%
%
Section~\ref{sec:avd-wetten} and~\ref{sec:avd-curvature} have discussed two ways to improve gradient dynamics thin-film models (\ref{eq:onefield:gov}) by amending the free energy functional. In particular, they discussed how to extract wetting potentials from microscopic models and how to employ exact expressions for the energy instead of their long-wave approximations. The present section considers the mobility function $Q(h)$. In the context of hydrodynamic thin-film equations obtained via long-wave approximations from the Navier-Stokes equations \cite{OrDB1997rmp}, the mainly discussed expressions are $Q\sim h^3$ (no slip at the solid substrate), $Q\sim h^2 (h+l_s)$ (weak slip with slip length $l_s$ at the substrate) and $Q\sim h^2$ (intermediate slip) \cite{MuWW2005jem}.

A further extension of the interpretation and application of thin-film models arises when considering the ultrathin adsorption layer (sometimes called ``precursor film'') that in the case of partially wetting at equilibrium coexists with macroscopic drops \cite{Genn1985rmp}. It is also found in front of advancing and behind receding contact lines \cite{PODC2012jpm}. Where the adsorption layer is concerned, the field $h$ in Eq.~(\ref{eq:onefield:gov}) should not anymore be strictly interpreted as ``film height'' but rather as ``adsorption'', i.e., as the mean number of molecules per substrate area.  Here, ``mean'' refers to an average over microscopic time scales.
This naturally allows one to describe situations where the density in the liquid film is not constant anymore (e.g., layering effects close to the substrate) or where there is no closed liquid film on the substrate but rather individual diffusing molecules \cite{HuTA2017jcp}. When the film is thicker the adsorption is proportional to the film height (the connecting factor is the constant equilibrium density of the liquid), i.e., there is a natural continuity between the two measures. This implies that even film heights, e.g., of equilibrium adsorption layers, smaller than a molecular diameter are acceptable if interpreted as adsorption.

Having this interpretation in mind, one has to request that for $h\to0$ the governing equation becomes the standard diffusion equation, that itself is also of gradient dynamics form (see section~II.B.1 of Ref.~\cite{ThAP2016prf}).  It turns out that this case corresponds to a constant mobility $Q$. This implies, a mobility function $Q\sim h^3 + C$ (where $C$ is an appropriate constant related to the wetting potential and the diffusion constant, see section~IV of Ref.~\cite{YSTA2017pre}) automatically switches between diffusion and convective motion when going from the adsorption layer to a drop or thick liquid film. Together with wetting potentials from section~\ref{sec:avd-wetten} that do not diverge for $h\to 0$, such a mobility allows to describe the spreading of drops with finite length adsorption layers (precursor films) and, in general, the advancement of contact lines over truly dry substrates. A first study of the influence of $C$ can be found in Ref.~\cite{YSTA2017pre}.

The incorporation of diffusion into thin-film models is also discussed in Refs.~\cite{YoPi2005pre,PODC2012jpm,HLHT2015l}. A piecewise mobility function that switches between diffusion and convection at a critical height is introduced to describe droplet motion due to surface freezing and melting \cite{YoPi2005pre}.  Ref.~\cite{PODC2012jpm} discusses convective and diffusive advancement in the context of adiabatic and diffusive films close to the contact line region using, however, a diverging wetting energy and standard cubic hydrodynamic mobility. This results in a height-dependent diffusion coefficient that diverges for $h\to0$ (see their section 3.2.2.B). Finally, Ref.~\cite{HLHT2015l} considers the redistribution stage of deposition experiments where organic molecules are deposited on a solid substrate with heterogeneous wettability. It is discussed that different instability modes of ridgelike transients lead to different nonlinear evolution pathways. It is further indicated that different transport mechanisms during the redistribution only affect the ratio of the time scales of the individual process phases but not the sequence of the observed morphologies. The particular cases studied are convective transport without slip ($Q\sim h^3$), transport via diffusion in the film bulk ($Q\sim h$) and via diffusion at the film surface ($Q\sim C$). The direct combination of the different transport channels into the same mobility function proposed in Ref.~\cite{YSTA2017pre} should also be seen in the context of the ongoing discussion of the mechanisms of contact line motion \cite{Blak2006jcis,QiWS2006jfm,PiEg2008pre,BEIM2009rmp,XuQi2010jcp,SaKa2011el,Pome2011epjt,PODC2012jpm}.

In the future, the direct combination of different transport channels within a single gradient dynamics evolution equation for film height/adsorption should be studied more systematically employing wetting potentials extracted from microscopic models. Also, transport coefficients and mobility functions extracted from microscopic models should be passed to the mesoscale models transforming qualitative comparisons as, e.g., between kinetic Monte Carlo and thin-film simulations \cite{TBHT2017jcp}, into quantitative ones. An alternative way of incorporating diffusion-like processes into mesoscopic hydrodynamics are stochastic thin-film equations \cite{GrMR2006jsp,FRSJ2007prl}. There it is shown that an incorporation of noise also results in a relative change of the dewetting time scales.
%
%%%%%%%%%%%%%%%%%%%%%%%%%%%%%%%%%%%%%%%%%%%%%%%%%%%%%%%%%%%%%%%%%%%%%%%%%%%%%%%
\section{Gradient dynamics for complex liquids}\label{sec:avd-grad}
%%%%%%%%%%%%%%%%%%%%%%%%%%%%%%%%%%%%%%%%%%%%%%%%%%%%%%%%%%%%%%%%%%%%%%%%%%%%%%%
%
In section \ref{sec:intro} we have presented the thin-film (long-wave) equation for simple liquids on horizontal substrates (\ref{eq:film}) as gradient dynamics (\ref{eq:onefield:gov}) and the appendix shows how this form can be derived from Onsager's principle.  In principle, it should also be possible to bring all long-wave models into such a form that are derived in the overdamped (or Stokes-) limit for complex fluids in relaxational situations, i.e., without additional energy sources or sinks. For thin films of nematic liquid crystals this is done, e.g., in Ref.~\cite{MeCO2009jpm,LCAK2013pf,LKTC2013jfm,LaCK2018jfm} for situations where the nematic director field is enslaved to the film height profile (for cases of weak and strong anchoring at the interfaces) and the evolution equation is again Eq.~(\ref{eq:onefield:gov}) with appropriate energy functional and mobility.

It is less known that this is also possible for most long-wave models that involve more than one field, e.g., for two-layer films or films of solutions, suspensions and mixtures, i.e., hydrodynamic thin-film models reviewed in sections II.C and VI of Ref.~\cite{CrMa2009rmp}, and models in Refs.~\cite{WaCM2003jcis,NaTh2010n,BMBK2011epje,FrAT2012sm,KaRi2014jfm}.  Once the gradient dynamics form is established one can develop a plethora of thin-film models for certain classes of complex fluids by amending the energy functionals along the lines discussed in section~\ref{sec:avd-curvature}.

Generalising the derivation of conserved, nonconserved and mixed gradient dynamics models given in the appendix, close to equilibrium one can write a general thin-film system that is characterised by the set of $n$ scalar fields $\vecg{\psi} = (\psi_1,\psi_2,..., \psi_n)=(\{\psi_a\})$ (e.g., layer thicknesses, local surfactant or solute amount, adsorption or precipitation of solute at substrate), as $n$ coupled evolution equations in gradient dynamics form
\begin{equation}
 \partial_t \psi_a=\nabla\cdot \left[\sum_{b=1}^n Q^\mathrm{c}_{ab}
   \nabla \frac{\delta \mathcal{F}}{\delta \psi_b}\right]
- \sum_{b=1}^n Q^\mathrm{nc}_{ab} \frac{\delta \mathcal{F}}{\delta \psi_b}
\label{eq:nn3}
\end{equation}
combining conserved and nonconserved dynamics that are both governed by the same underlying energy functional $\mathcal{F}[\vecg{\psi}]$. Each field can have a conserved and/or nonconserved dynamics as encoded in the respective mobility matrices $Q^\mathrm{c}_{ab}(\vecg{\psi})$ and $Q^\mathrm{nc}_{ab}(\vecg{\psi})$ that are $n \times n$ dimensional, positive definite and symmetric. For more background and the three-field example of a thin-film model for a liquid layer or drop covered by a soluble surfactant see \cite{ThAP2016prf}.

In particular, the gradient dynamics approach of Eq.~(\ref{eq:nn3}) was
applied to several situations that can be described by two scalar fields. This includes
the dewetting of two-layer films \cite{PBMT2004pre,PBMT2005jcp,JHKP2013sjam,HJKP2015jem},
coupled decomposition and dewetting processes for a film of binary
mixture \cite{Thie2011epjst,ThTL2013prl} (without Marangoni forces,
see corresponding discussion in section IV.B of \cite{ThAP2016prf})  
and the evolution of a liquid film covered by insoluble surfactant
\cite{ThAP2012pf} (for soluble surfactant see \cite{ThAP2016prf}). 
In the case without evaporation or other nonconserved processes, all these two-field models can be brought into the form
\begin{eqnarray}
\partial_t h \,&=&\,
\vec{\nabla}\cdot\left[Q_{hh}\vec{\nabla}\frac{\delta \mathcal{F}}{\delta h}\,+\,Q_{h\psi}\vec{\nabla}\frac{\delta \mathcal{F}}{\delta \psi}\right]
\nonumber\\
\partial_t \psi \,&=&\,
\vec{\nabla}\cdot\left[Q_{\psi h}\vec{\nabla}\frac{\delta \mathcal{F}}{\delta
                      h}\,+\,Q_{\psi \psi}\vec{\nabla}\frac{\delta
                      \mathcal{F}}{\delta \psi}\right]
\label{eq:twofield}
\end{eqnarray}
where $h$ and $\psi$ are two fields with conserved dynamics. The
mobility matrix is 
\begin{equation}
\tens{Q}\,=\,\left( 
\begin{array}{cc}  
Q_{hh} & Q_{h\psi} \\[.3ex]
Q_{\psi h} &Q_{\psi \psi}
\end{array}
\right)
\,=\,\frac{1}{\eta}\left( 
\begin{array}{cc}  
h^3/3 & a\,h^2\psi \\[.3ex]
a\,h^2\psi \phantom{xx}& b\,h\psi^2+ c(\psi) \psi
\end{array}
\right),\\[-.5ex]
\label{eq:twolay-mob}
\end{equation}
where $\eta$ is a viscosity, and the respective energy functionals
$\mathcal{F}$ have a clear physical meaning and may be obtained via
the coarse-graining procedures of statistical physics. Normally, they
contain entropic and/or interaction terms as well as terms penalising
strong field gradients (see
Refs.~\cite{PBMT2004pre,JHKP2013sjam,Thie2011epjst,ThTL2013prl,ThAP2012pf,ThAP2016prf}).

The particular mobilities are as follows: For the two-layer film of Ref.~\cite{JHKP2013sjam}, $h$ and $\psi$ correspond to the thickness of the lower and upper layer, respectively, and in (\ref{eq:twolay-mob}) $a=1/2$, $b=1$ and $c(\psi)=\eta\psi^2/3\mu$ where $\eta$ and $\mu$ are the viscosities of lower and upper layer, respectively. For the film of binary mixture of Refs.~\cite{Thie2011epjst,ThTL2013prl}, $h$ and $\psi$ correspond to the thickness of the film and the effective solute layer height $\psi=h\phi$ (where $\phi$ is the height-averaged concentration), respectively, and in (\ref{eq:twolay-mob}) $a=b=1/3$ and $c(\psi)=\widetilde D$ is the diffusive mobility of the solute.  For the film with insoluble surfactant of Ref.~\cite{ThAP2012pf}, $h$ and $\psi$ correspond to the thickness of the film and the surfactant concentration projected onto the cartesian plane, and in (\ref{eq:twolay-mob}) $a=1/2$, $b=1$ and $c(\psi)=\widetilde D$ is the diffusive mobility of the surfactant. Note, that Ref.~\cite{XuTQ2015jpcm} employs Onsager's variational principle \cite{Doi2013} to develop a solvent-solute symmetric mixture model (without surface activity) that is also valid at high solute concentrations. It has a mobility matrix that is cubic in the fields as (\ref{eq:twolay-mob}) but also symmetric with respect to an exchange of the two fields (local amounts of solvent and solute, respectively). There exist other two-field gradient dynamics models, e.g., for membrane dynamics \cite{SaGo2007pre,HiKA2012pre} or as DDFTs for mixtures \cite{Arch2005jpcm,ArRT2010pre}.

It has two advantages to bring hydrodynamic thin-film models into the form (\ref{eq:twofield}) [or (\ref{eq:nn3})]: First, it is a relatively simple way to show that a particular long-wave approximation still gives a thermodynamically consistent model. For instance, such models with specific energies are found in Refs.~\cite{NaTh2010n,KoGF2009el,KGFC2010l}. In consequence, it also indicates that many literature models are incomplete as they incorporate ad hoc amendments into hydrodynamic long-wave equations, e.g., by introducing concentration-dependent Derjaguin pressures for solutions and suspensions or by using nonlinear equations of state for surfactants.  This is avoided through the second advantage of the form (\ref{eq:twofield}), namely, that amending the energy functionals $\mathcal{F}$ while keeping the mobilities (also cf.~section~\ref{sec:avd-curvature}) automatically results in thermodynamically consistent evolution equations for the coupled fields.  For instance, changing the local energy of the surfactant from purely entropic to a combination of entroic and interaction terms \cite{ThAP2012pf}, results in a nonlinear equation of state (i.e., nonlinear Marangoni flux) \textit{and} non-Fickean diffusion (and more complex adsorption/desorption dynamics in the case of soluble surfactant \cite{ThAP2016prf}).

For the present special volume on complex wetting it is of particular interest that the amendments of the energy functionals can consist in the introduction of wetting energies that do not only depend on film height but also on other fields. For instance, a concentration-dependent wetting potential gives a concentration-dependent Derjaguin pressure and also results in an additional Marangoni-type flux, affects diffusion, evaporation, and adsorption/desorption. The exploration of the possibilities of these gradient-dynamics models has only just started \cite{ThTL2013prl,STBT2015sm,SaTB2016jcp,TSTJ2018l}. Future directions could include investigations of the coupling of wettability with structural phase transitions of surfactants or solutes, and studies of terrassed spreading of drops of nanosuspensions.

Beside changing the energies, one can also amend the mobilities, e.g., by incorporating slip at the substrate as done for two-layer systems in \cite{MuWW2005jem} or by allowing for solvent diffusion along the substrate as discussed for single layers in Refs.~\cite{HLHT2015l,YSTA2017pre} and section~\ref{sec:avd-mobility}.  Another important question relevant, e.g., for particle suspensions, is the dependence of the liquid viscosity on solute concentration. This can be easily incorporated, as long as the liquid is Newtonian. There are few thin-film models for non-Newtonian liquids, e.g.\ Refs.~\cite{FlKi2004jem,BeGr2005jpcm,MWRB2006epje,AfMW2007pre}, and to our knowledge none was discussed as gradient dynamics. It would also be interesting to incorporate surface viscosity \cite{SDDR2010el} that would most likely only affect the mobility matrix.
%
%%%%%%%%%%%%%%%%%%%%%%%%%%%%%%%%%%%%%%%%%%%%%%%%%%%%%%%%%%%%%%%%%%%%%%%%%%%%%%%
\section{Thin-film biofilm models}\label{sec:avd-biofilm}
%%%%%%%%%%%%%%%%%%%%%%%%%%%%%%%%%%%%%%%%%%%%%%%%%%%%%%%%%%%%%%%%%%%%%%%%%%%%%%%
%
The previous sections have discussed recently proposed improvements for the description of the behaviour of drops, films or single contact lines of simple and complex liquids on solid substrates. These improvements are closely interlinked with the ability to bring the various hydrodynamic thin-film equations into the gradient dynamics form of Eqs.~(\ref{eq:nn3}) [or (\ref{eq:onefield:gov}) or (\ref{eq:twofield})]. In consequence, one might assume that the resulting models are only able to describe relaxational behaviour, i.e., the evolution towards local or global equilibria. This is, however, not the case. Instead, the gradient dynamics models can serve as a ``thermodynamically consistent core'' that is then supplemented by selected out-of-equilibrium driving terms. For instance, an addition of appropriate potential energies to the energy functional allows one to investigate sliding droplets on an incline \cite{EWGT2016prf}. Including ``comoving frame terms'' gives out-of-equilibrium models for the transfer of a liquid film (dip coating) or surfactant layer (Langmuir-Blodgett transfer) from a liquid bath onto a withdrawing moving substrate \cite{WTGK2015mmnp}.  Furthermore, additional in- and out-fluxes of material \cite{BeMe2006prl,TDGR2014l} or energy \cite{PBMT2005jcp} can be incorporated that break the gradient dynamics structure.  This approach allows one to use all improvements discussed in sections~\ref{sec:avd-wetten} to~\ref{sec:avd-grad} also in many thin-film models for out-of-equilibrium processes. A tentative classification of such nongradient additions to gradient-dynamics evolutions equations for scalar order parameter fields is given in \cite{EGUW2018springer}.

Here, we particularly highlight the recent example of hydrodynamic biofilm models that are developed to describe the interplay of biological and physicochemical processes in the early stages of the spreading of bacterial colonies \cite{WASB2011mb} on moist agar substrate in an air atmosphere \cite{FPBV2012sm,SAWV2012pnasusa,DeML2014if,WaWH2015cmmm}.
To consistently describe the interplay of wettability, capillarity and osmotic fluxes on the one hand and biological growth on the other hand, Refs.~\cite{TrJT2016ams,TJLT2017prl} employ a two-field model as Eq.~(\ref{eq:twofield}) with (\ref{eq:twolay-mob}), where the two fields represent the aqueous nutrient medium and biomass (bacteria and bacteria-produced extracellular matrix), and incorporate osmotic fluxes by adding nonconserved terms as in Eq.~(\ref{eq:nn3}).  This thermodynamically consistent model is then driven out of equilibrium by the proliferation of biomass that itself triggers an osmotic influx.  Also here, models with more complex physics may be developed following the pathways laid out in section~\ref{sec:avd-wetten} to~\ref{sec:avd-grad}.  See, for instance, models that incorporate the dynamics of bacteria-produced bio-surfactants \cite{ARKW2009pnasusa,FPBV2012sm,TrJT2018sm}.

In another example, insoluble self-propelled particles swimming on a liquid film are described via a combination of a standard thin-film model for a surfactant-covered thin film with terms resulting from active swimming \cite{AlMi2009pre,PoTS2016epje,PoTS2016springer}. To go beyond the considered limit of dilute self-propelled particles towards higher concentrations one could follow the path of amendments possible within the gradient structure of the passive part of the model.
%

%%%%%%%%%%%%%%%%%%%%%%%%%%%%%%%%%%%%%%%%%%%%%%%%%%%%%%%%%%%%%%%%%%%%%%%%%%%%%%%
\section{Conclusions}
\label{sec:conc}
%%%%%%%%%%%%%%%%%%%%%%%%%%%%%%%%%%%%%%%%%%%%%%%%%%%%%%%%%%%%%%%%%%%%%%%%%%%%%%%
%
The present contribution to the volume on complex wetting has highlighted and commented on important developments of the past five years that widen the scope and reach of mesoscopic thin-film (or long-wave) hydrodynamic models employed to describe the dynamics of thin films, drops and contact lines of simple and complex liquids on solid substrates.  We have focused on developments that are based on the reformulation of various mesoscopic thin-film hydrodynamic models as gradient dynamics on underlying energy functionals. We have briefly presented the general approach, followed by discussions how to improve these models by amending the underlying energy functional and mobility functions, how to obtain gradient dynamics models for some complex liquids, and how to incorporate processes beyond relaxational dynamics.

We have argued on the one hand that the possibility of such a reformulation is a good measure of the quality of a model that describes a relaxational dynamics. However, at points this poses interesting general questions as an asymptotically correct model might not have the form of a gradient dynamics and vice versa. We believe that the gradient dynamics form of governing equations is the ``higher value''. This implies then that procedures of asymptotic expansions (e.g., the application of the long-wave approximation) should be applied in a way that at each order the asymptotic model represents an exact gradient dynamics.

On the other hand we have laid out that the gradient dynamics form facilitates the widening of the scope and reach of mesoscopic thin-film hydrodynamics in several ways. In particular, we have briefly discussed in section~\ref{sec:avd-wetten} how to extract wetting potentials without divergence problems for vanishing film height from microscopic models, pointing out that similar methodologies still need to be developed to extract transport coefficients. In section ~\ref{sec:avd-curvature} we have proposed to improve lowest order models by improving the energy functional within the gradient dynamics formulation. For the considered interface-dominated systems this has a larger impact on the model behaviour than the mobilities. Up to now this has only be done systematically for a few systems. Section~\ref{sec:avd-mobility} has then focused on a possible extension of the mobility beyond hydrodynamic motion to capture the transition between diffusive and convective transport occurring when modelling drops on ultrathin films or adsorption layers. The subsequent section~\ref{sec:avd-grad} has discussed how to obtain gradient dynamics models for complex liquids and has detailed some of the options for two-field models, in particular, for two-layer films and films of solutions, suspensions and mixtures. Here, we express the hope that for these more complex models such a presentation in gradient dynamics form will become a standard as it allows for an easy comparison and validation of models. Finally, section~\ref{sec:avd-biofilm} has ventured away from relaxational dynamics and has argued that also most models of out-of-equilibrium processes in the discussed system class should have a well defined gradient dynamics core supplemented by nonequilibrium driving terms. This has been illustrated at the example of recent models for osmotically spreading biofilms. 
Each section has presented and commented on the respective main argument from the recent literature and has discussed pertinent open questions. Naturally, there have been other interesting developments in dynamic wetting that have not been covered here as they are not directly related to the gradient-dynamics approach. Examples are the development of path-continuation techniques for the determination of families of self-similar solutions related to film rupture \cite{TsBT2013ijam,DTZF2017n,DFTK2018prl}, the development of time-stepping  techniques to study solutions of the thin-film equation that have an unknown support \cite{Pesc2015jcp}, studies of the fully nonlinear dynamics of stochastic thin-film dewetting \cite{NCMK2015pre} (cf.~\cite{MeRa2005jpcm,GrMR2006jsp}), studies of (de)wetting dynamics through the study of invariant manifolds of a suitable dynamical system \cite{TsGT2014epje,BeGK2016nama}, and the derivation of Smoluchowski-type statistical models for the ensemble dynamics of sliding drops from single-drop bifurcation diagrams \cite{WTEG2017prl}.

\acknowledgments The mentioned work by the author was supported through grants by the European Union (grant PITN-GA-2008-214919 (MULTIFLOW)); the Deutsche Forschungsgemeinschaft (DFG, Transregional Collaborative Research Center TRR 61 (B2), project TH 781/8-1); the German-Israeli Foundation for Scientific Research and Development (Grant No. I-1361-401.10/2016); and the DAAD (Campus France, PHC PROCOPE Grant No.~35488SJ). We also acknowledge funding of the wider field by the DFG through Priority Program 2171 ``Dynamic Wetting of
Flexible, Adaptive, and Switchable Substrates'' and by the Banff International Research Station for Mathematical
Innovation and Discovery (BIRS) through the Workshop ``Modelling of Thin Liquid Films-Asymptotic Approach vs.\ Gradient Dynamics'' (19w5148, 28.4.-3.5.2019).
The author also expresses his gratitude for the frequent discussions with collaborators and colleagues on gradient dynamics and closely related questions. This includes, in particular, A.J. Archer, A.v.\ Borries Lopes, L.J. Cummings, T. Frohoff-H\"ulsmann, S.V. Gurevich, S. Engelnkemper, A.L. Hazel, A. Heuer, K. John, E. Knobloch, M.H. K\"opf, L. Kondic, T.-S. Lin, D. Peschka, L.M. Pismen, M. Plapp, T. Qian, M. Shearer, J.H. Snoeijer, H. Stark, F. Stegemerten, W. Tewes, D. Todorova, S. Trinschek, H. Uecker and M. Wilczek.

\appendix
\section{Deriving basic gradient-dynamics via Onsager's principle}
\label{sec:grad-dyn}
In the main part of this work we have used the thin-film equation in
the form (\ref{eq:onefield:gov}) of a gradient dynamics for a single
scalar, conserved order parameter field.  Here, we briefly indicate
how to derive this general form using Onsager's variational principle
\cite{Doi2011jpcm,Doi2013}, i.e., by minimising the Rayleighian with
respect to appropriate fluxes. The Rayleighian corresponds to the sum
of the total time-derivative of the energy functional and the
dissipation functional. Our derivation is similar to the derivation of
the Cahn-Hilliard equation in \cite{Doi2013}. Note that after the conserved case we also consider nonconserved and mixed case.
\subsection{Conserved order parameter}
For a conserved order parameter field $\psi(\vec{r},t)$, the
time-evolution equation must have the form of a continuity equation (conservation law), i.e., 
\begin{equation}
\frac{\partial \psi}{\partial t} = -\nabla \cdot\vec{j}_\mathrm{c}
\label{eq:cop1}
\end{equation}
where $\vec{j}_\mathrm{c}$ is the $\psi$-conserving flux. The total time derivative of the general energy functional $F[\psi]$ is
 \begin{equation}
\frac{dF}{d t} = \int d\vec{r} \frac{\delta F}{\delta\psi}\frac{\partial \psi}{\partial t}
\label{eq:cop2}
\end{equation}
and the simplest dissipation functional is quadratic in the flux
\begin{equation}
\Phi = \frac{1}{2}\,\int d\vec{r}\,\zeta _\mathrm{c}\, |\vec{j}_\mathrm{c}|^2
\label{eq:cop3}
\end{equation}
where the friction $\zeta _\mathrm{c}\ge0 $ may still depend on $\psi$.
We collect the expressions and obtain the Rayleighian $R=dF/dt + D$ as
\begin{equation}
R = \int dx\, \left[\tfrac{1}{2}\zeta _\mathrm{c} |\vec{j}_\mathrm{c}|^2+ 
\frac{\delta F}{\delta\psi}\frac{\partial \psi}{\partial t}
\right] = 
\int dx\, \left[\tfrac{1}{2}\zeta _\mathrm{c} |\vec{j}_\mathrm{c}|^2+ 
\left(\nabla\frac{\delta F}{\delta\psi}\right) \cdot\vec{j}_\mathrm{c}
\right]
\label{eq:cop4}
\end{equation}
where in the final step we used (\ref{eq:cop1}) and
  integrated by parts.
We now minimize $R$, i.e., determine $\delta R/\delta
\vec{j}_\mathrm{c}=0$, and obtain the flux
\begin{equation}
\vec{j}_\mathrm{c}= -\frac{1}{\zeta _\mathrm{c}}\, \nabla\,\frac{\delta F}{\delta\psi}
\label{eq:cop5}
\end{equation}
With continuity (\ref{eq:cop1}), we finally obtain the evolution equation
    for a single conserved scalar order
    parameter field $\psi$. 
\begin{equation}
\partial_t\,\psi\,=\,\vec{\nabla}\cdot \left\{Q_\mathrm{c}(\psi)\, \vec{\nabla}\,
\frac{\delta F[\psi]}{\delta \psi}\right\} 
\label{eq:de}
\end{equation}
with the energy functional $F[\psi]$ and mobility function
$Q_\mathrm{c}(\psi)=1/\zeta _\mathrm{c}\ge0$. This corresponds to Eq.~(\ref{eq:onefield:gov}).
Note, that the present derivation only gives the general form. The particular mobility function $Q(\psi)$
needs to be obtained through a long-wave expansion of the full
governing equations (also cf.~\cite{XuTQ2015jpcm}).
\subsection{Nonconserved order parameter}
For a single nonconserved order parameter field $\psi$, no continuity equation is needed. Instead the time evolution
is directly given by a nonconserving flux (strictly speaking a creation/deposition/condensation rate)
\begin{equation}
\frac{\partial \psi}{\partial t} = j_\mathrm{nc}
\label{eq:ncop1}
\end{equation}
Eq.~(\ref{eq:cop2}) is still valid and the simplest dissipation functional reads
\begin{equation}
\Phi = \frac{1}{2}\,\int dx\,\zeta_\mathrm{nc} \, j_\mathrm{nc}^2
\label{eq:ncop2}
\end{equation}
Variation of the Rayleighian w.r.t.\ $ j_\mathrm{nc}$ directly gives
\begin{equation}
\partial_t\,\psi\,=\,- Q_\mathrm{nc} \frac{\delta F[\psi]}{\delta \psi}
\label{eq:gdg}
\end{equation}
with the mobility $Q_\mathrm{nc}(\psi)=1/\zeta_\mathrm{nc}\ge0$ 
\subsection{Mixed conserved and nonconserved dynamics}
Finally, a system can have mixed conserved and nonconserved dynamics. 
The general form is the balance equation
 \begin{equation}
\frac{\partial \psi}{\partial t} = -\nabla\cdot\vec{j}_\mathrm{c}  + j_\mathrm{nc}.
\label{eq:mixop1}
\end{equation}
Eq.~(\ref{eq:cop2}) is still valid, but the dissipation functional has two contributions
\begin{equation}
\Phi = \frac{1}{2}\,\int d\vec{r}\,\left(\zeta_\mathrm{c} \,
  |\vec{j}_\mathrm{c}|^2 + \zeta_\mathrm{nc} \,
  j_\mathrm{nc}^2 \right).
\label{eq:mixop2}
\end{equation}
As a result, the variation of the Rayleighian w.r.t.\ $\vec{j}_\mathrm{c}$ and $j_\mathrm{nc}$,
gives after  use of (\ref{eq:mixop1}) the mixed gradient dynamics
\begin{equation}
\partial_t\,\psi\,=\, \vec{\nabla}\cdot \left(Q_\mathrm{c}(\psi)\, \vec{\nabla}\,
\frac{\delta F[\psi]}{\delta \psi}\right) - Q_\mathrm{nc} \frac{\delta F[\psi]}{\delta \psi}
\label{eq:mixevol}
\end{equation}
with mobilities as above. For a generalisation to $n$ coupled scalar order parameter
  fields see section~\ref{sec:avd-grad}.

%ADAPT DIRECTORY
% \bibliography{%
% $HOME/Home/Bibliography/uwelitall,%$
% $HOME/Home/Bibliography/books} %$
%\bibliography{Thie2018final} %$
%
% TO PRODUCE PAPER bib FILE: CREATE aux FILE, THEN RUN
% bibtool -i $HOME/Home/Bibliography/uwelitall.bib -i $HOME/Home/Bibliography/books.bib -x Thie2018final.aux -o Thie2018final.bib

\end{document}